\documentclass[twocolumn,superscriptaddress]{revtex4-1}
\usepackage{graphicx}
\usepackage{caption}
\setcitestyle{super}
\usepackage{subcaption}
\usepackage{comment}
\usepackage{amsmath}
\usepackage{amssymb}
\usepackage{dcolumn}
\usepackage{bm}

\usepackage{url}

\begin{document}

\title{Fixed-node errors in real space quantum Monte Carlo at high densities:  closed-shell atomic correlation energies 
}

\author{Lubos Mitas}

\affiliation{
 Department of Physics, North Carolina State University, Raleigh, North Carolina 27695-8202, USA\\
}

\date{\today}

\begin{abstract}
We consider non-relativistic electron correlation energies of heavy noble gas atoms including the superheavy element Og. The corresponding data enables us to quantify fixed-node errors in real space quantum Monte Carlo methods as a function of the atomic number $Z$. 
We confirm that single-reference trial function nodes lead to an overall trend of mild decrease in recovered correlation energy with the increasing $Z$. This agrees with our previous study that has shown increasing fixed-node biases with the increasing electron density. We also estimate the value of the linear term in the asymptotic expansion of the atomic correlation energy.
\end{abstract}

%

\maketitle

\section{Introduction}
Quantum Monte Carlo (QMC) methods belong to the most advanced approaches for capturing many-body effects in electron-ion systems. In the real space of particle coordinates, QMC methods are based on explicitly correlated wave functions and computationally feasible techniques for many-body solutions of the stationary Schr\"odinger equation. Practical QMC calculations rely on the so-called fixed-node/phase (FN/FP) approximations in order to avoid the well-known fundamental obstacle and corresponding inefficiencies from fermion signs \cite{sandrobook, qmcrev,lucasrev,kolorenc2011}. The fixed-node/phase
approximation impacts the outcomes and introduces a certain level of the so-called fixed-node/phase biases.   These biases are typically small and real space QMC methods that include variational Monte Carlo (VMC) and  FN diffusion Monte Carlo (FNDMC) enable us to calculate highly accurate total energies and energy differences for hundreds of valence electrons \cite{kolorenc2011, can, cody, gani}.
 Indeed, QMC methods are particularly well-positioned for large-scale calculations due to their favorable scaling in the number of particles so that calculations of condensed systems and periodic solids have become routine \cite{kent_qmcpack_2020}.

An established measure of method accuracy for solving many-body eigenvalues and eigenstates 
is the electron-electron correlation energy defined as the difference between the exact and Hartree-Fock energies. 
Recently, Nakano co-workers \cite{Nakano2020} have presented fixed-node calculations of heavy noble gas atoms, including the superheavy element Oganesson (Og), $Z=118$. 
Interestingly, in the original presentation and analysis \cite{Nakano2020}
  the fixed-node biases appeared to be somewhat larger than would be expected from the past studies \cite{qmcrev,lucasrev,kolorenc2011,sandrobook,gani,umrigar1993,needsrev, rasch2014}. For example, for the Xe atom the single-reference fixed-node DMC appeared to recover $\approx$ 78\%, while for Og only $\approx$  71\% of the correlation energy. In addition, 
geminal-based wave functions that employ 
pairs instead of single-particle orbitals did not substantially change this picture  \cite{Nakano2020}. 
These points are of particular interest, since previous studies have indicated that an increase in the electronic density is indeed related to the increase of  fixed-node biases \cite{rasch2014}. 

Another relevant analysis of atomic correlation energies has been carried out within the Density Functional Theory (DFT) \cite{Burke2016,Burke2018, Burke2020, Burke2024}. The focus has been on the overall description of correlation as a function of atomic number $Z$ as well as on improved versions of practically used approximate functionals. 
Some of this analysis has broader ramifications. The asymptotic terms found in heavy atoms are related to the corresponding terms in high-density jellium \cite{Loos2016} which is an important  paradigmatic model for condensed matter systems.

Our present work is focused on shedding more light on the correlation energy in heavy atoms as well as on more precise study of relationship of fixed-node errors and electron density in the limit of large atomic numbers. 
For this purpose, we 
first consider estimations
of correlation energies for noble gas heavy atoms using the best available data from correlated wave function calculations based on basis set expansions. 
Although
our focus will be on non-relativistic energies, 
we will be using both relativistic and non-relativistic calculations for estimating desired values. In particular, we obtain the correlation energy of the Og atom using an extrapolation of previously calculated 
data from many-body perturbation theory (MBPT) and Coupled Cluster (CC) approaches. (We recall that in relativistic calculations
 the correlation energy is given by the difference between the lowest energy of the determinant of one-particle spinors 
 and the exact value of the corresponding relativistic Hamiltonian \cite{Mani2021}.) 
We also carry out FNDMC calculations for the Rn atom which was absent in the previous QMC studies so as to make the corresponding data for noble gas atoms complete. 
Interestingly, we find that the fixed-node errors are smaller than approximately estimated in the mentioned study \cite{Nakano2020} due to the fact that their reference (ie, estimated exact) correlation energies appear to be a bit too large, roughly by 10\%.
At the same time, we indeed confirm that 
the FNDMC bias increases with the density as we demonstrated for lighter atoms and molecules some time ago \cite{rasch2014}.  Having collected this data, we also analyze the asymptotic behavior of the atomic correlation energies in the limit of large atomic numbers to complement recent DFT
studies \cite{Burke2016,Burke2018}. Finally, we point out implications of our findings for accurate QMC calculations
where changes of fixed-node errors with electron density can have a significant impact,
for example, for systems under pressure. 

\section{Correlation energies of noble gas atoms }

 {\bf Og atom correlation energy.} For the purpose of this study, we estimate the all-electron Og atom correlation energy using data calculated by many-body methods based on basis set expansions. In particular, 
highly accurate values of correlation energies up to Rn were obtained some time ago \cite{Thakkar2011} within non-relativistic framework using
 Coupled Cluster (CC) method with  extensive basis sets. These CC estimations have included also extrapolations to complete basis set limits and therefore it essentially established exact values for lighter elements within $\approx$ 1 mHa accuracy or better.
 For heavier elements these results are close to the exact ones on relative scale, with systematic uncertainty estimated to be about 1-3 \%, up  to Rn ($Z=86$). Very recently, an independent study presented calculations of several heavy atoms including group-18 elements using relativistic many-body perturbation theory (MBPT) as well as relativistic CC with comparably extensive basis sets \cite{Mani2021}. These calculations include also a few closed-shell superheavy elements such as Og ($Z=118$), from the closed-shell inert/noble gas atom group. The most extensive calculations  employed basis sets with angular momentum channels up to $\ell=9$ \cite{Mani2021}. 

\begin{figure}[ht]
\includegraphics[width=3.5in,clip]{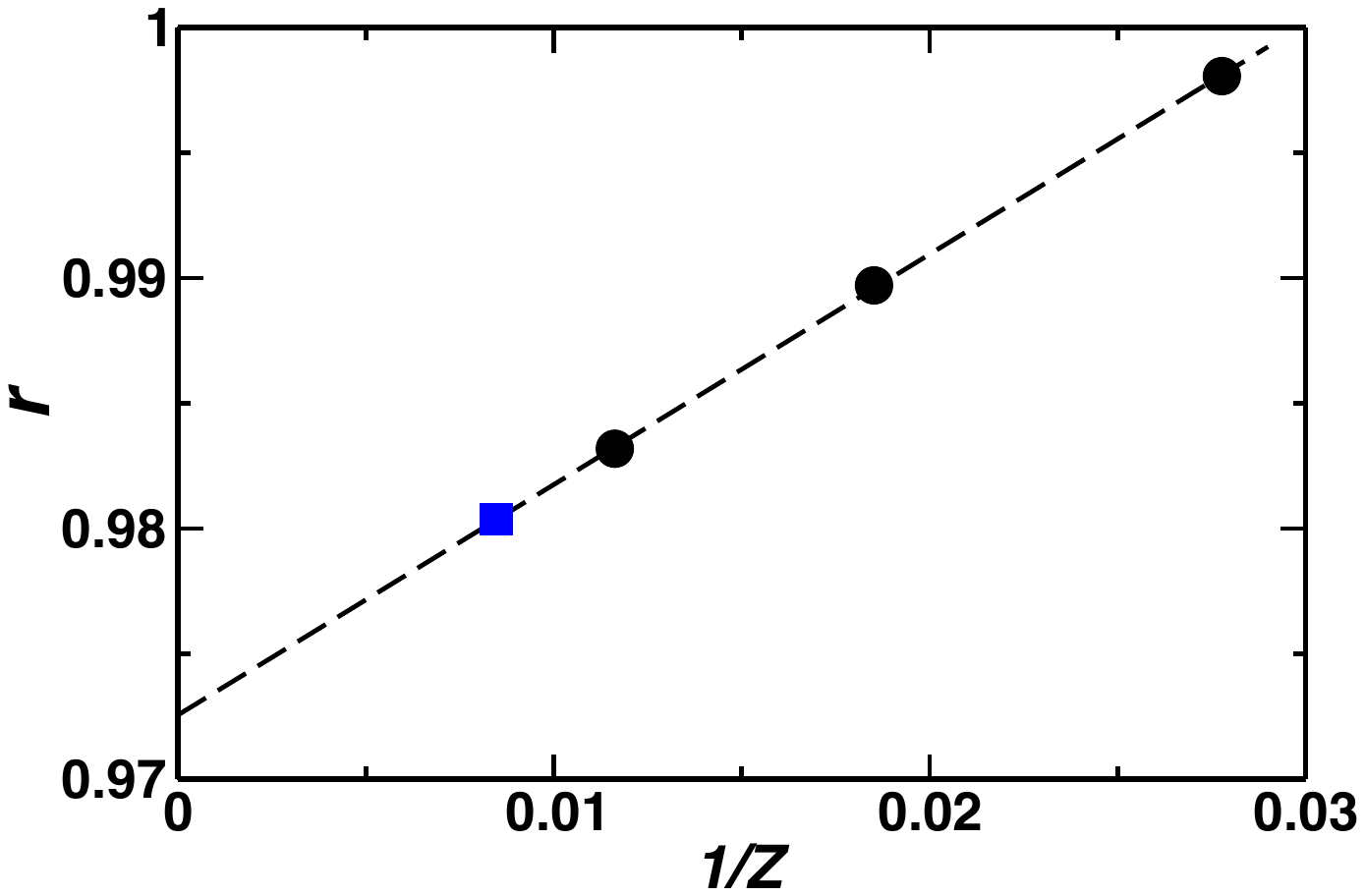} 
\caption{ Ratio of CC (non-relativistic) and MBPT (relativistic) atomic correlation energies for Kr, Xe and Rn (circles) as a function of $1/Z$. The dashed line is a linear fit that is used to estimate the CC value for Og (square), see Tab. I. 
}
\label{fig:ratio}
\end{figure}

After some analysis we found that the data from these two studies show very similar and systematic behavior. 
We list the most systematic results for correlation energies from both mentioned papers, see Table \ref{tab:tab1}. In particular, we show non-relativistic CC correlation energies that include extrapolations to the complete basis set limit \cite{Thakkar2011}. Similarly, we include the most extensive MBPT relativistic calculations \cite{Mani2021}. Although these two sets of calculations were done independently and in different settings, they show very remarkable consistency that is clear to see once we evaluate the ratio of these two sets of correlation energies.
For the heaviest inert elements, this ratio shows strikingly small variation
and almost perfect linear dependence on $1/Z$, see Table \ref{tab:tab1} and Fig.\ref{fig:ratio}.
Note that the range for this ratio  for $Z=36-118$ is very small, within $\approx$ 2\%. This is much smaller than, for example, ratio of MP2 and CC energies or other many-body estimators that exhibit much less systematic behavior, see, for example, Ref. \cite{Thakkar2012}.
A linear fit of 
this ratio for Kr, Xe, and Rn 
as a function of $1/Z$
is shown in Fig.\ref{fig:ratio} and in Tab. \ref{tab:tab1}. From this fit we estimate the ratio for $Z=118$ which then leads to an estimate of the non-relativistic correlation energy of Og, as given in the Tab. \ref{tab:tab1}. (We tested also extrapolation based only on the two heaviest atoms, Xe and Rn. That has resulted in almost identical estimate for Og, within about 1.5 mHa, Table \ref{tab:tab1}.) The original publication \cite{Thakkar2011} estimates systematic accuracy of the CC values to be 
about 1.9\%, 1.2\%, 1.4\%
and 3.0\% for Ar, Kr, Xe and Rn, respectively. The sizes of basis sets used by Mani and co-workers \cite{Mani2021} have been comparably extensive so that we assume similar level of uncertainty of 
their MBPT results. 
We therefore estimate the accuracy of our value  $\approx$ 3\% which is adequate for our subsequent analysis below since we will be looking at the biases which are significantly larger.  
This value therefore completes the non-relativistic set of correlation energies for the inert ( noble gas) atoms with accuracy that is approximately consistent with lighter elements estimations as given in Ref.\cite{Thakkar2011}.

\begin{table*}[!htbp]
\centering
\caption{Calculated  correlation energies (mHa)  
of heavy atoms from MBPT (relativistic)\cite{Mani2021} and CC (non-relativistic)\cite{Thakkar2011} methods together with corresponding ratios.  Value of $r$ for Og (with asterisk) is found from linear extrapolation of ratios for Kr, Xe and Rn as a function of $1/Z$, Fig.1. 
Estimated value of $E_{\rm corr}^{CC}$ for Og (with asterisk) is obtained as $E_{\rm corr}^{CC} = r E_{\rm corr}^{MBPT}$. Ref. \cite{Thakkar2011} estimates systematic
uncertainties of the CC values  for heavier atoms as 
1.2\%, 1.4\% and 3.0\% for 
Kr, Xe and Rn, respectively. We assume similar $\approx$ 3\% uncertainty for the Og atom. 
}
\label{tab:tab1}
\smallskip
\begin{tabular}{llllll}
\hline
\hline
atom $Z$ & $-E_{\rm corr}^{MBPT}$  &  
 $-E_{\rm corr}^{CC}$ &  $r$=$E_{\rm corr}^{CC}/E_{\rm corr}^{MBPT}$  \\
\hline
Kr 36   &1853.2   & 1849.6 & 0.998057   \\
Xe 54    & 3031.4 & 3000.2  & 0.989708 \\
Rn 86  & 5619.5 &  5525.0       & 0.983184 \\
Og 118   & 8910.9 & 8735.8$^*$ &  0.980350$^*$ & \\
\hline
Kr, Xe, Rn: $r=c_0+c_1/Z$, & 
$c_0=0.972554$ & $c_1=0.919919$ & \\
Xe, Rn: $r=c_0+c_1/Z$, &  $c_0=0.972175$ & $c_1=0.946796$, &$\to$ Og:  $-E_{\rm corr}=$ 8734.5 \\
\hline
\hline
\end{tabular}
\end{table*}

{\bf QMC data.}
Let us now turn to analysis of FNDMC  biases for noble gas atoms calculated with single-reference trial functions. 
Although in these systems with some effort one can increase the node accuracy besides the single-reference and therefore decrease the fixed-node biases, we opt for the single-reference FNDMC that provides an essential baseline for 
QMC studies in general.

\begin{figure}[ht]
\includegraphics[width=3.5in,clip]{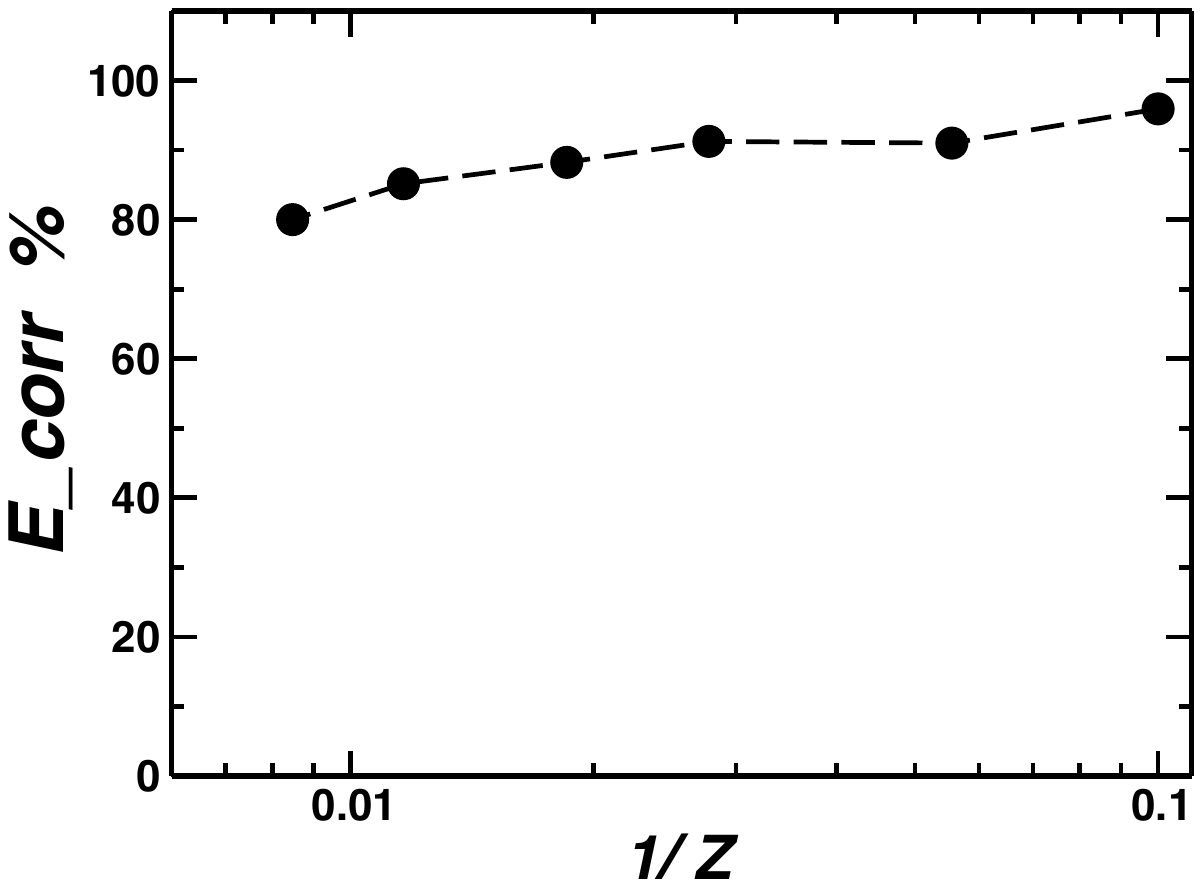} 
\label{ratio}
\caption{Obtained $E_{corr}$ (in \%) by fixed-node DMC with single-reference trial function for Ne, Ar, Kr, Xe, Rn and Og atoms.  Uncertainties vary from fraction of a percent for lighter atoms to approximately the size of the symbol for Rn and Og. The line is for guiding the eyes. 
}
\end{figure}

We note that the noble gas atoms are favorable for single-reference studies due to simplicity of closed shells and qualitatively similar electronic properties appropriate for analysis of asymptotic trends.
In QMC most of these elements have been already calculated \cite{ma2005b,Nakano2020} with the exception of the Rn atom.

\begin{table*}[!htbp]
\centering
\caption{
Fixed-node DMC correlation energies (mHa) and their percentages of exact values for noble gas atoms using CC data and  estimate from above for the Og atom. We report  
previous DMC calculations \cite{ma2005b} with single-reference Slater-Jastrow (SJ) and with AGP-Jastrow (AGPJ) trial functions
\cite{Nakano2020}. 
Our calculations for Xe and Rn are based on single-reference Slater-Jastrow trial functions. Error bars (in parenthesis) below the last shown digit are dropped.
}
\label{tab:be_interact}
\smallskip

\begin{tabular}{lrrllrrl}
\hline
\hline
atom $Z$ & $-E_{\rm corr}$   & DMC & &  Ref. &  \qquad DMC& & Ref. \\
 & \qquad (est.$^*$) &\qquad\qquad SJ& & & \qquad AGPJ & & \\
\hline
Ne 10  & 392 &  376 & 95.9 & \cite{ma2005b} & 379 & 96.7 & \cite{Nakano2020} \\
Ar 18 & 733 & 667 & 91.0 & \cite{ma2005b} & 679 &92.6 & \cite{Nakano2020} \\
Kr 36  &  1850  & 1688 & 91.2 &\cite{ma2005b} & 1716 &92.8 & \cite{Nakano2020}\\
 &  & 1691(7) & 91.4(4) &  this work &  & & \\
Xe 54   & 3000  &2647 & 88.2  & \cite{ma2005b} & 2697 & 89.9 & \cite{Nakano2020}\\
Rn 86 & 5525 & \qquad 4634(52)  &83.9(9)            &  this work \\
Og 118  & 8736$^*$ & &  & & 7082 &81.1 & \cite{Nakano2020}\\
\hline
\hline
\end{tabular}
\end{table*}

In order to fill this gap, we have carried out QMC of Rn as well as Xe atom with the single-reference  Slater-Jastrow trial wave functions.
The Jastrow term contains functions that depend on inter-particle distances, in particular, they describe electron-electron cusp and related two-electron correlations. The orbitals for the Slater determinant were obtained from numerical Hartree-Fock solver \cite{Shirleyphd1991,Koto1997} and we reproduced the previously obtained Hartree-Fock energy \cite{Saito2009} within a fraction of mHa. In actual VMC and DMC calculations the radial parts of orbitals were represented by cubic splines with fine radial step 
that was adequate to reproduce the HF energy in the corresponding variational Monte Carlo check. Since we were not focused on high overall precision, we used conventional \cite{qmcrev} fixed-node DMC algorithm which is rather inefficient  for cases with rapidly varying electronic densities (in particular, around nuclei in heavier atoms \cite{umrigar93}). However, the resulting error bars are acceptable for present purposes since the systematic uncertainties of calculated and estimated exact correlation energies are still significantly larger. 

\begin{figure}[ht]
\includegraphics[width=3.4in,clip]{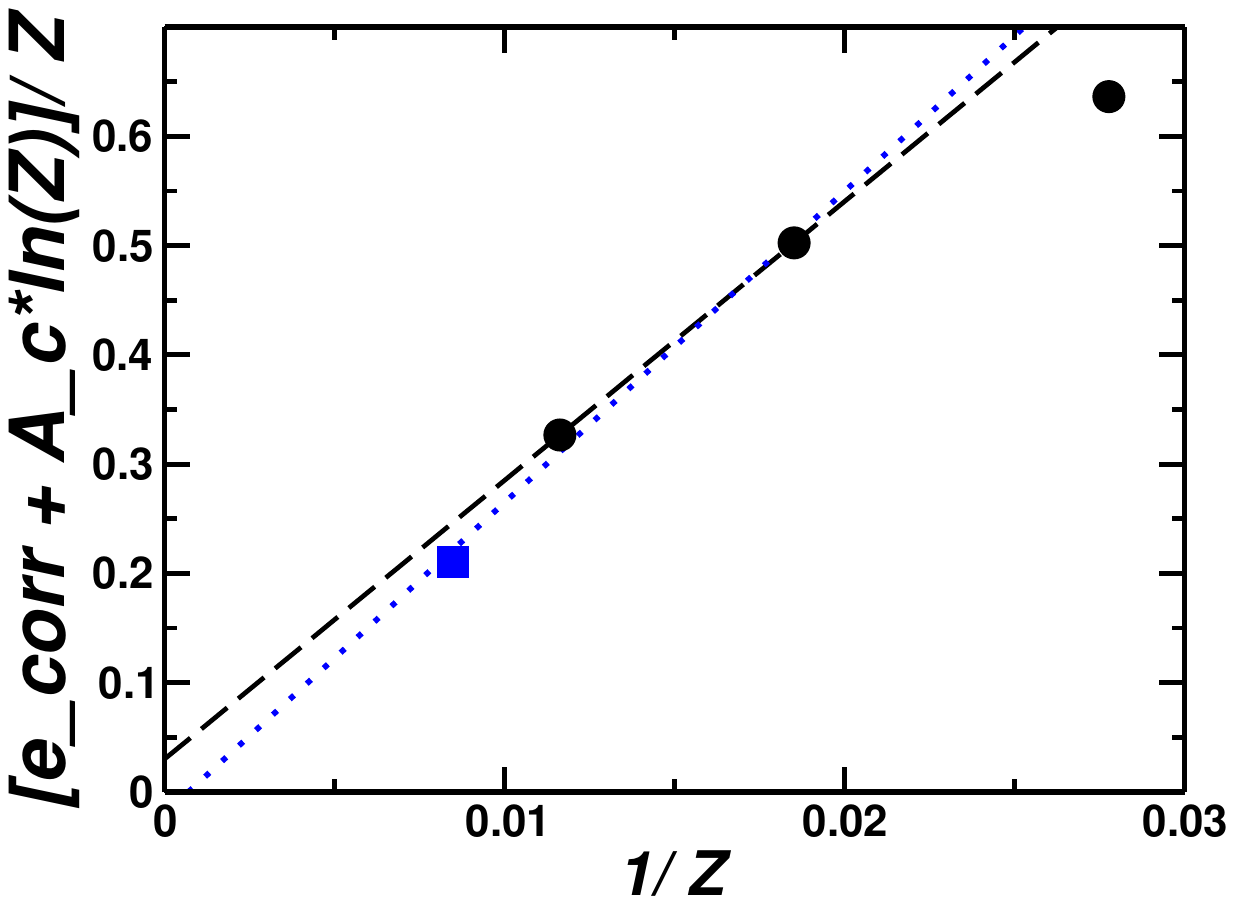} 
\caption{Plot of $(e_{corr}+A_C\ln(Z))/Z$  as a function of $1/Z$ for atoms Kr, Xe and Rn  (circles) and Og (square). The dashed linear fit includes only Xe and Rn points, the dotted one includes also the estimated data point for Og.
In this plot the coefficient $B_C$ in Eq. \ref{eq:ecor} is given by the slope and its values for the plotted fits are $\approx$ 26 and $\approx$ 28 mHa.
Since the exact asymptote should pass through the origin, the impact of  non-asymptotic and shell-filling effects  appears to be present even in the heaviest atoms. This is visible also in nonzero but otherwise marginal absolute constants of the two  fits. Imposing zero absolute term gives only a 
minor change of the constant $B_C$, to about 27 mHa. 
}
\label{fig:cor}
\end{figure}

Using the results from references \cite{Nakano2020,ma2005b} we plot the percentage of the FNDMC correlation energies 
in Fig. \ref{fig:cor}.
We see overall decreasing trend as a function of $1/Z$ with a mild deviation from the trend for Ar. Due to irregularities,
an approximate extrapolation suggests that in very large $Z$ limit one would recover roughly 75\% of the correlation energy. (It is understood that here we ignore the well-known instabilities, ie, the electronic instability of neutral atoms and also the nuclear instability, both appearing beyond values of $Z \approx 120$ or so.) 

\section{Extrapolation and asymptotic terms}

As shown previously, see for example recent Refs. \cite{Burke2016,Burke2018}, the non-relativistic correlation energy of atoms in the limit of $Z\to\infty$ can be expressed as 
\begin{equation}
e_{corr}=E_{corr}/Z=- A_C\ln(Z)+B_C + ....
\label{eq:ecor}
\end{equation}
where the total atomic correlation energy $E_{corr}$ is defined as customary as $E_{corr}=E_{exact}-E_{HF}$
and the value of $A_C$ is known exactly\cite{Burke2016, Kunz2010}. It is related to the well-known expression for the correlation energy of the homogeneous electron gas and the corresponding constant is given by
$A_C=2c_0/3$ where $c_0=(1-\ln2)/\pi^2$ [Ha]. The correlation energy for jellium is expressed in terms of
the Wigner radius $r_s$ while in atoms this dependence is replaced by $1/Z$. It has been demonstrated that the linear-logarithmic term is asymptotically universal as shown also for
other Coulomb interactions models such as spherically or harmonically confined electron gas \cite{Loos2011,Loos2012,Loos2016}.
We carry out estimation of the $B_C$ using data for  Xe, Rn and Og by neglecting possible contributions from higher order terms and shell-filling effects. Clearly, even for  
these heaviest atoms we can expect some 
non-asymptotic effects and in order to assess this better,
 we plot the quantity $(E_{corr}+A_CZ\ln(Z))/Z^2$ as a function of $1/Z$. This plot indicates whether there is a trend to converge to the origin, since that should be the exact limit for  $1/Z\to 0$ and the value of $B_C$ is then given by the slope of the linear fit.
The data indeed show tendency to extrapolate to zero although the shell-filling effects are still significant even for Rn and presumably also for Og. In order to get some insight into the range of reasonable values and also to assess the sensitivity to the choices of data to be included, we tried two fits: the first one that uses only Xe and Rn data points, the second one with the three elements Xe, Rn and Og, see Fig. \ref{fig:cor}. The plot includes also  the value  for
Kr which appears off the overall trend and therefore it was not used in the fits. 
The resulting extrapolations cross the vertical axis at non-zero values, however, the asymptotic trends are clear and provide justification of the presented analysis. The range of nonzero offsets show the degree of uncertainty caused by the presence of the shell-filling effects and other non-asymptotic behavior. In these fits the  parameter $B_C$ varies rather marginally, the two obtained values are  $\approx$ 26 and 28 mHa. If we enforce the linear fit to pass through the origin we obtain $\approx$ 27 mHa. 
All of these values are somewhat smaller than the corresponding estimation by Burke et al\cite{Burke2016, Burke2018} of $B_C\approx 37$ mHa, that was derived by optimizing   expressions for exchange-correlation functionals while taking 
into account large number of elements in the periodic table. This difference is therefore understandable, since here we have avoided  lighter elements
and also open shell-effects. 
 Interestingly, our value appears to be closer to the RPA estimate of $B_C \approx 18 $ mHa that should be reasonably accurate at higher densities \cite{Burke2016, Burke2018}.

\section{Discussion and Conclusions}
The single-reference FNDMC method recovers substantial amount of correlation energy which, however, shows slow decreases with increasing atomic number and the corresponding electronic density.  For Og atom the amount is around 80\%, which is markedly better than the previously given estimate ($\approx$  70 \%) \cite{Nakano2020}. For practical  QMC electronic structure calculations of properties this is meaningful only as an asymptotic limit since valence-only related properties of ordinary condensed matter and molecular systems have much smaller densities and consequently exhibit much smaller errors.

It is interesting to consider, at least qualitatively, the origin of why high densities lead to larger fixed-node errors. 
The typical correlated trial functions describes the two-particle electron-electron and electron-ion/pseudoion cusps explicitly, however, their impact on the nodal surface is more complicated and it is also more difficult to capture.  Nakano  and co-workers \cite{Nakano2020} employed the BCS-like antisymmetrized geminal pair trial functions to probe this aspect. However, the improvements for this particular effect have been rather modest. Indeed, Tab. 2
shows that it is about 1.6\% for all calculated atoms. 
The node  deformations  require inclusion of excitations with
virtuals adapted to the core scales as combined with the excitations in valence space. Sufficiently accurate trial function would require very large numbers of core-core, core-valence and valence-valence excitations since the $3N-1$-dimensional node is clearly an all-degrees-of-freedom, collective property of the exact eigenstate.

The next point we want to emphasize is that the dependence of the fixed-node bias on density should be taken into consideration 
in calculations which include systematic electronic density changes,
most notably in calculations of condensed systems at high pressures. In such applications even small increase in the fixed-node error could affect the results very significantly, for example, the QMC equations of state (eg, total energy vs volume). 
For example, more steep increase of the total energy with decreasing volume would distort estimation of
transition pressures and phase changes towards higher values. The increasing complexity of the nodal surface with localized higher densities (in core or semicore regions) appears as a significant factor even for systems with $sp-$elements 
such as C, N, and O
\cite{rasch2014}. Clearly, these effects
require better benchmarking and trial function improvements that would eliminate or at least better delineate this type of bias. 

Let us summarize the key  findings. 

a) We have estimated the non-relativistic value of the atomic correlation energy for the Oganesson atom using  previous many-body wave function calculations
based on extensive basis set expansions and extrapolations. The estimated value of 8736 mHa $\pm$ 3\%  completes high accuracy non-relativistic data set for the group of noble elements based on Coupled Cluster method \cite{Thakkar2011}. 

b) We have carried out analysis of the fixed-node errors produced by single-reference trial functions in QMC diffusion Monte Carlo calculations.
 We see that the amount of recovered correlation energy,  in general, decreases with increasing atomic number $Z$ and with the corresponding increase of the electron density.
For the heaviest elements it is between 80 and 90\% which is roughly 10 \% less than in typical FNDMC calculations with lighter elements. Clearly, the present calculations are still not very precise and more accurate studies, both with basis set correlated wave function methods and QMC approaches, are highly desirable.  
The same applies to calculations of valence energy differences although is is expected
that these would be less impacted by deep core states so that one expects most of these errors to
cancel out. Again, the true extent of such cancellations in all-electron calculations for very heavy elements have yet to be studied. In effective core potential (ECP) calculations these difficulties are mostly avoided up to their fidelity to all-electron, fully correlated settings (see, for example, the new 
set of correlation consistent ECPs in Ref.\cite{haihan2024} and references therein). However,
even in valence-only calculations increases in electron density make nodal surfaces more complicated with some influence on resulting accuracy \cite{rasch2014}.

c) Finally,  we estimate 
the value of the linear term in the asymptotic expansion of the atomic correlation energy as a function of $Z$ for $Z\to \infty$. We find value of 26-28
mHa, which is somewhat smaller than 
previous estimates
\cite{Burke2016,Burke2018}. Considering that the data for the heaviest atoms still exhibit shell-filling effects and some degree of non-asymptotic behavior, further refined analysis might provide more precise estimation. In general,  the
overall trend in data indicates the presence of the linear-logarithmic term at high densities. This term is universal since, as explained previously, it is caused by the short range effects from the Coulomb singularity \cite{Loos2011,Loos2016,Burke2016}.   

Our study therefore maps out the behavior of single-reference QMC calculations at high electronic densities and points out possible impacts on calculations of electronic properties that can be influenced by this type of fixed-node biases. It also suggests that it is worth to consider ideas for addressing these shortcomings by better adapted  trial wave function constructions and as well as by more extensive study how the valence energy differences could be affected.

\section{Acknowledgments}

This work has been supported by the NSF grant DMR-2316007.
This study, as well as several previous ones, have been inspired by papers and ideas of Sandro Sorella and his collaborators.

\bibliography{main}

\end{document}